\RequirePackage{fix-cm}
\documentclass{svjour3}                     
\smartqed  
\usepackage{graphicx}
\usepackage{multirow}
\usepackage[caption=false]{subfig}
\usepackage{comment}
\usepackage{soul}
\usepackage{booktabs}
\usepackage{color}
\usepackage{amsfonts}

\usepackage{xcolor} 
\usepackage{amssymb}

\newcommand{\shakshi}[1]{{\color{black}{\textbf{}}\color{black}{1}}{\color{black}}}

\newcommand{\rev}[1]{{\color{black}{#1}}}

\title{GAME-ON: Graph Attention Network Based Multimodal Fusion For Fake News Detection
}
\begin{document}


\author{Mudit Dhawan$^*$\thanks{$^*$Work done while at IIIT Delhi} \and Shakshi Sharma$^{**}$\thanks{$^{**}$Mudit Dhawan \& Shakshi Sharma have equal contribution}    \and  Aditya Kadam    \and
        Rajesh Sharma  \and Ponnurangam Kumaraguru
}



\institute{Mudit Dhawan \at Microsoft Research, India\\
              \email{t-mdhawan@microsoft.com} \and
              Shakshi Sharma \at Bennett University, Greater Noida, India\\
              \email{shakshi.sharma268@gmail.com} \and  Rajesh Sharma \at   Institute of Computer Science, University of Tartu, Estonia \\
              \email{rajesh.sharma@ut.ee} \and Aditya Kadam \and Ponnurangam Kumaraguru \at
              Department Of Computer Science, IIIT Hyderabad, India \\
              \email{aditya.kadam@research.iiit.ac.in, pk.guru@iiit.ac.in}
              }

\date{Received: date / Accepted: date}

\maketitle

\begin{abstract}  
Fake news being spread on social media platforms has a disruptive and damaging impact on our lives. Multimedia content improves the visibility of posts more than text data but is also being used for creating fake news. Previous multimodal works have tried to address the problem of modeling heterogeneous modalities in identifying fake news. However, these works have the following limitations: (1) inefficient encoding of inter-modal relations by utilizing a simple concatenation operator on the modalities at a later stage in a model, which might result in information loss; (2) training very deep neural networks with a disproportionate number of parameters on small multimodal datasets result in higher chances of overfitting. To address these limitations, we propose GAME-ON, a Graph Neural Network based end-to-end trainable framework that allows granular interactions within and across different modalities to learn more robust data representations for multimodal fake news detection. We use two publicly available fake news datasets, Twitter and Weibo, for evaluations. GAME-ON outperforms on Twitter by an average of 11\% and achieves state-of-the-art performance on Weibo while using 91\% fewer parameters than the best comparable state-of-the-art baseline. For deployment in real-world applications, GAME-ON can be used as a lightweight model (less memory and latency requirements), which makes it more feasible than previous state-of-the-art models.

\end{abstract}

\section{Introduction}\label{sec:introduction}

The fast growth of social media has created a perfect environment for the diffusion of information, be it genuine or fake. However, without any quality control over the disseminated information, fake news has far-reaching consequences \cite{zhao2015enquiring}. 
\textcolor{black}{For example, the influence of fake news during the 2016 presidential election in the United States \cite{bovet2019influence}, the dissemination of numerous myths, and misleading information about the COVID-19 pandemic \cite{melki2021mitigating,sharma2021identifying}. Furthermore, posts that utilize visual cues such as images and videos entice more users to join the debate and engage with the content. } 
Fake news developers use the tactics of adding visual information to the text to craft more appealing posts to deceive users \cite{verstraete2021identifying}.
Therefore, detecting fake news while taking into account multimodal data is of utmost importance.

\begin{figure}[t]
    \centering
    \includegraphics[width=1\linewidth]{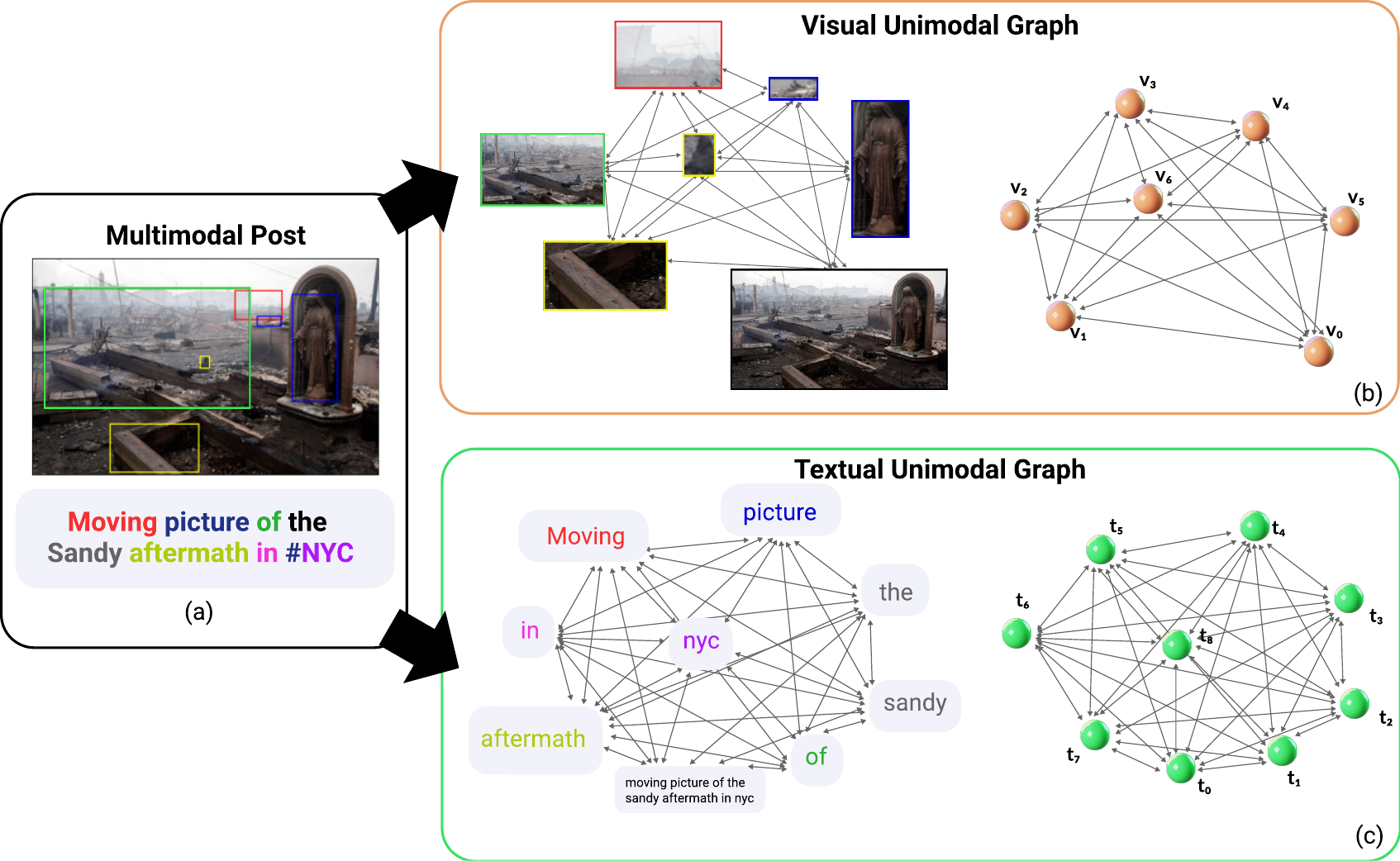}
    \caption{Overview of the graph construction pipeline for the GAME-ON  framework. \emph{(a)} Given a multimodal post (news sample), taken from the Twitter dataset, we extract individual fully-connected graphs for both the modalities. \emph{(b)} We find objects from the image and extract their feature representations ${v_i}$. \emph{(c)} For the textual graph, we first tokenize the text and extract their feature representations ${t_i}$.}
    \label{fig:graph_example}
\end{figure}

\textcolor{black}{
Recent works focus on fake news detection tasks broadly from two perspectives: unimodal (text or visual) and multimodal (text and visual) paradigms. There have been significant works in unimodal settings, such as using the extracted emotions of the posts \cite{zhang2021mining} and writing styles of the articles \cite{potthast2017stylometric} to train deep learning based models. In addition, advances from Knowledge-graph based techniques have also been employed \cite{nikopensius2023reinforcement,mayank2022deap}. In addition, metadata information such as investigating the role of message creation time in the early identification of fake news \cite{ramezani2019news} has also been utilized. Not only content but social information such as comments and reposts \cite{kwon2013prominent,sanh2019distilbert,shu2017fake,singhal2020spotfake+}, and user reputation \cite{yu2017convolutional} are also taken into account in detecting fake news.
}

Recently, there has been growing interest among researchers in the field of multimodal fake news detection.
Various deep-learning based architectures have been proposed
\textcolor{black}{, such as a Bi-Variational Autoencoder model 
\cite{khattar2019mvae,wang2018eann} to extract the shared representations of multimodal data, an event-discriminator module in the Event Adversarial Neural Network (EANN) architecture \cite{wang2018eann}, to make the model event-invariant using adversarial style learning.}
In addition, transfer learning strategies have become increasingly popular in identifying fake news \cite{singhal2019spotfake,singhal2020spotfakeplus}.

Researchers have also focused on intra- and inter-modal interactions by fusing different modalities using self and cross-attention networks at the cost of unnecessarily large models \cite{wu2021multimodal,qian2021hierarchical}.
\textcolor{black}{The authors of \cite{wu2021multimodal} introduced a multimodal co-attention networks model that takes care of intermodal interactions by fusing different modalities using co-attention maps at the cost of a complex model. }

The shortcoming of the previous works is the inefficient fusion of different modalities using complex models. Fusing modalities using simple concatenation at a later point (after trainable parameters or fine-tuned specialized encoders)

in the model, in particular, can result in information loss. In addition, previous works that utilize the concatenation operator for encoding the intermodal relationship fail to explicitly address the \emph{heterogeneity gap} \cite{peng2019cm} that arises in multimodal data.

Even studies that have attempted to address the aforementioned issues utilized a complex model with a large number of parameters \cite{qian2021hierarchical}, possibly leading to higher chances of overfitting.

In this work, we propose an end-to-end trainable graph neural network framework for detecting news
veracity. To enhance the performance, we employ early stage fusion of heterogeneous modalities to prevent information loss. Our framework achieves promising results on the two widely and publicly used datasets in the fake news domain.
Further, to save computational costs, we focused on reducing the model's complexity, i.e., lowering the model's size. We show that our model uses significantly fewer parameters than the baselines.

\textcolor{black}{Wang et. al. \cite{wang2022instance} encodes a news item as a heterogeneous multimodal graph, but unlike ours, they fix the number of nodes which limits the information that could be extracted from the news item and might lead to information loss. Specifically, they employ separate modality-specific models to encode graph features, which doesn't bridge the heterogeneity gap and also utilize distinct stages and parameters (model blocks) for applying inter- and intra-modal fusion. We found this work to be the closest to  our work.} 
\rev{

\textbf{The motivation behind using Graphs:}
\textcolor{black}{
We consider a multimodal post with both textual and visual content and create a multimodal graph as is done in a previous work \cite{wang2022instance} (see Figure \ref{fig:graph_example}). However, unlike its graph with a single type of feature vector for each modality, we extract and create two types of nodes- global and fine-grained feature representation nodes for each modality}.  We include both semantic (text as a whole, and image as whole ) and fine-grained (syntactic/word-level, and object-level) representations for both text (Figure \ref{fig:graph_example} \emph{(c)}) and image  (Figure \ref{fig:graph_example} \emph{(b)}) modalities, respectively.
These different types of representations for each modality (\textit{nodes}) help the model to learn complex relationships (\textit{edges}) within and across modalities in data more efficiently in a \textit{graphical} manner. 
}

Graph Neural Networks (GNNs) have revolutionized many fields, including network science, visual dialogue, and have achieved excellent performance on numerous tasks \cite{chen2020hgmf,han2020graph,mai2020analyzing}. Also, in contemporary multimodal representation learning works, only a few have employed these powerful GNN techniques. 
\textcolor{black}{However, these works either introduce tensor factorization-based methods that are sensitive to outliers or utilize separate stages for inter- and intra-modal encoding. Therefore, unlike our proposed framework, the former introduces unwanted complexity, and the latter can not simultaneously model inter- and intra-modal relationships.}
Building on the gaps from the previous works and the recent success of GNNs, the main contributions of our work are as follows:

    \noindent (1). We propose \textbf{GAME-ON}, a novel end-to-end trainable GNN-based framework\footnote{https://github.com/mudit-dhawan/GAME-ON} for identifying fake news on social media platforms using multimodal data; 
    
    \noindent (2). The proposed framework allows for granular interactions across (inter)- and within (intra)- modalities to fuse them early in the framework, decreasing information loss; 
   
    \noindent (3). With fewer trainable parameters, we propose a simple approach when compared to complicated state-of-the-art models. 
    In particular, our model has $\sim$\textbf{91\%} fewer parameters than the best comparable baseline; and 
    
    \noindent (4). We assess our model on two publicly available real-world datasets, MediaEval 2015 and Sina Weibo.
    Our model outperforms the current state-of-the-art models on Twitter by an average of \textbf{11\%}, and achieves competitive performance on Weibo, with significantly fewer parameters.

The rest of the paper is organized as follows. Section \ref{sec:related} examines the literature. The proposed framework is discussed in Section \ref{sec:method}. Experiments using real-world datasets are covered in Section \ref{sec:experiments}. Finally, Section \ref{sec:conclusion} discusses the implications and future work.

\section{Related Work}\label{sec:related}

\subsection{Unimodal Fake News Detection}
There has been a variety of research that focuses on one modality in fake news detection tasks, such as text or images.
Specifically, text features, such as analyzing the emotions of posts \cite{zhang2021mining,sharma2023misinformation} and article writing styles \cite{potthast2017stylometric,mayank2022deap,sharma2022facov,sharma2022mis}, are being utilized for training deep learning based models. Furthermore, metadata has been utilized in research to identify fake news. One such study \cite{ramezani2019news} looks at the role of message creation time in early fake news detection. 

Besides content, social aspects such as comments and reposts \cite{kwon2013prominent,sanh2019distilbert,shu2017fake,singhal2020spotfake+}, as well as user reputation \cite{yu2017convolutional}, are being considered. In images, the authors primarily used rich visual information with distinct pixel domains and utilized a multi-domain visual neural network to detect fake news \cite{qi2019exploiting,jagtap2021misinformation}.
Meta-learning, a novel paradigm for domain shift datasets, has also been utilized to detect fake news in the Syrian war dataset, which can be applied to other armed conflicts related datasets \cite{salem2021meta}. 

Another study \cite{nan2021mdfend} highlights the challenges of domain shift in single-domain detection models. In particular, the authors built a fake news detection dataset with nine domains and then proposed a multi-domain fake news detection model that uses a domain gate to aggregate numerous representations generated by a mix of experts in a multi-domain environment.

\subsection{Multimodal Fake News Detection}\label{sec:multimodal}

In the past, many approaches \cite{butt2021goes,castillo2011information,sharma2021identifying,shu2017fake} focused primarily on extracting unimodal features from the data to detect fake news. 
However, research has shown that the inclusion of multimodal features improves model performance \cite{chen2020hgmf}.
A multimodal fake news detection is made up of two separate encoders for text and visual modalities. Researchers have attempted to add either a novel fusion approach or a sub-task to aid in the fusion \cite{khattar2019mvae,wu2021multimodal}.

Various deep-learning based architectures have been used in detecting fake news on multimodal data \cite{khattar2019mvae,wang2018eann,zhou2020safe}. 
\textcolor{black} {A Bi-Variational Autoencoder based architecture was proposed to extract the shared representation of multimodal data, that is, learn correlations across the tweets’ text and images extracted using the Bi-LSTM network and VGG-19, respectively \cite{khattar2019mvae}.
In another work, an event-discriminator module in the EANN architecture was introduced to remove event-specific features using adversarial style learning \cite{wang2018eann}. 
For textual and visual features extraction, Text-CNN and VGG-19 networks were used respectively and then concatenated to find a multimodal representation.
\cite{zhou2020safe} focused on the relationship between the visual and textual information in a news article and introduced a new Similarity-Aware fake news detection. The feature extraction methods employed in this work were different as they first converted the image into text using a pre-trained image-to-sequence model. They then used cosine similarity to capture the cross-modal similarity of two types of textual data.}
Other works \cite{singhal2020spotfakeplus,singhal2019spotfake} leveraged pre-trained transformer-based textual encoders and VGG-19 for image feature extraction to classify fake news via transfer learning.
Cross-attention networks are also heavily used in recent works that help in improving the accuracy of the models \cite{wu2021cross,ying2021multi}.
The shortcoming of the previously mentioned works is the inefficient encoding of inter-modal relations at a later stage in a model and the use of simple concatenation operators on the modalities, which could result in information loss.

Moreover, the existing works mentioned above lack an efficient fusion of multimodal features \cite{wang2020fake}. However, the authors in \cite{wu2021multimodal} presented a multimodal co-attention networks model in their fake news detection work, which takes care of inter modalities relations by fusing multiple modalities using co-attention maps at the cost of having a complex model. Peng et. al. \cite{peng2019cm} illustrate why it is necessary and challenging to bridge the heterogeneity gap caused due to inconsistent distributions and representations of various modalities. 

\rev{In the light of the recent widespread use of transformer-based models to extract semantic embeddings for both text and visual modalities, multiple works have been proposed to make both inference and fine-tuning using transformers more efficient \cite{zhou2023multi}. These works are complementary to our current work and can be used to further optimize the inference and training time for our pipeline. For efficient fine-tuning of these unimodal models, one such notable work is the Adapter-based fine-tuning of transformers \cite{pfeiffer2020adapterhub}. Adapter-based methods are used to efficiently finetune large encoder-based language models on a downstream task using smaller datasets. These methods add a small bottleneck layer after each transformer layer in these large models and only train these smaller layers which is efficient. However, these methods won’t be able to take into account the inter- and intra-modality fusion required for correctly classifying a multimodal piece of information as nuanced as an image article or tweet. These techniques are used majorly for unimodal tasks and haven’t been fully explored for settings that require heterogeneous sources of information such as text, images, etc at once. 
}

\noindent\textbf{The difference between GAME-ON and the previous multimodal transformer methods} is that the previous multimodal transformer methods have not taken into account fine-grained representations for both image and text modalities. Furthermore, most of these methods also utilize separate transformer blocks to learn intra-modal and inter-modal features. These distinct transformer blocks (adds to the computation cost of the pipeline) take into account a single global vector representation for image and text as inputs, thereby restricting the granular interaction between the modalities. 

\subsection{Graph Neural Networks (GNNs)}\label{sec:multimodal-gnn}
GNNs have excelled in many disciplines, including  network science, semantic forensics, health, visual dialogue \cite{mai2020analyzing,chen2020hgmf,han2020graph,monti1902fake,bian2020rumor,zhang2021multi,yin2020novel,gao2020multi,zhou2023multi}.

In non-multimodal settings, the authors used preference modeling to combine the pattern-based and fact-based textual models into one framework \cite{sheng2021integrating}. They employed a heterogeneous dynamic graph convolutional network to build preference maps, which they subsequently used to steer joint learning of pattern and fact-based models for final prediction.

Another work \cite{ni2021mvan} utilized graph attention networks utilizing propagation structure along with the users' tweets to detect fake news on two real-world Twitter datasets (Twitter 15 and Twitter 16), outperforming the state-of-the-art models. 
One application of GNNs include detect fake content in sustainable vehicular social networks \cite{guo2022mixed} wherein information security is a primary concern. This approach integrates both local and global semantics of the text to enhance feature space.
Another application of GNNs includes GraphSage wherein authors \cite{li2021meet} trained adversarial networks to minimize the adversarial risk on rumor detection tasks by leveraging objective facts from Wikipedia articles and subjective facts from social media posts.

In a multimodal environment, 
\cite{jiang2020kbgn} employed a graph to create fine-grained cross-modal semantic links between visual and text knowledge, as well as an adaptive information selection mode to get the relevant information.

\section{Methodology}\label{sec:method}

\begin{figure*}[t]
    \centering
    \includegraphics[width=\textwidth]{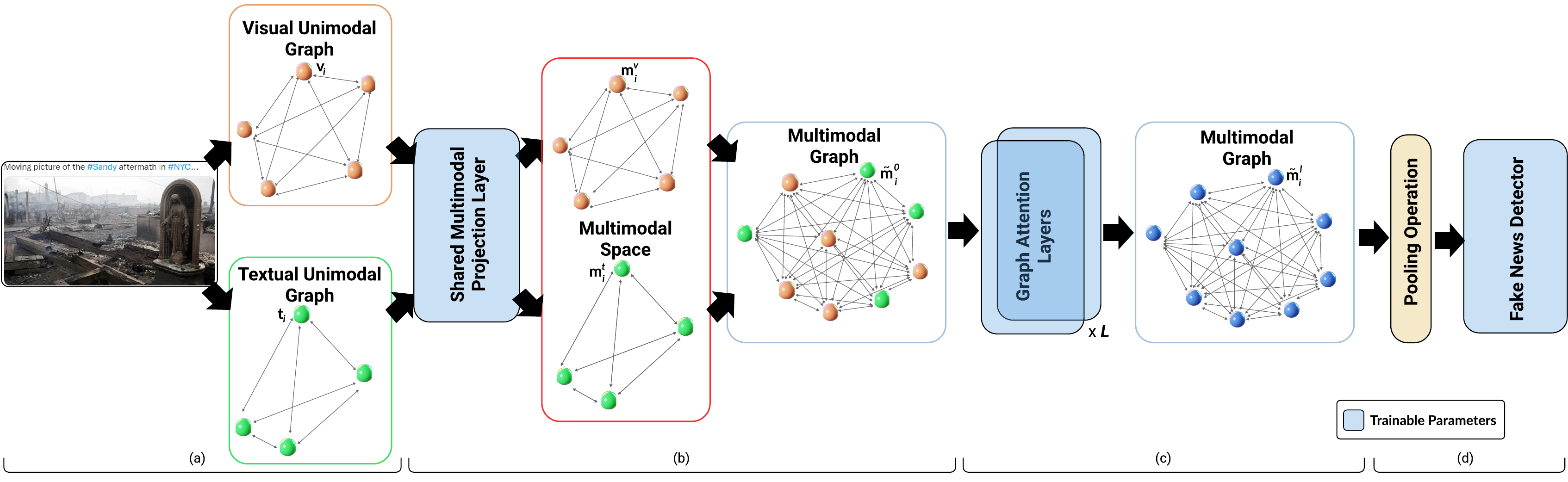}
    \caption{Overview of the GAME-ON framework. It consists of four stages.  \emph{(a)} Each modality is transformed into a unimodal visual and textual graph  (subsection \ref{subsec:feature-encoders}). \emph{(b)} \textcolor{black}{To establish common space representations of each modality,} 
    Both graphs are routed into fully-connected layers and inter-modality connections are introduced (subsection \ref{subsec:multimodal-space-graph}). \emph{(c)} \textcolor{black}{The multimodal }Graph is given to the graph attention layer (\emph{L}=1) to learn resilient representations  (subsection \ref{subsec:graph-layers}). \emph{(d)} The pooling and classification layer \textcolor{black}{are utilized to determine whether or not a news sample is fake}
    (subsection \ref{subsec:fake-news-detector}).}
    \label{fig:model_diagram}
\end{figure*}

\textbf{Model Overview.} The task of multimodal fake news detection can be described as a binary classification task, where the aim is to model the news (${N}$) sample's text(s) ${T}$ and image(s) ${V}$ to predict whether the given news is real or fake.\footnote{Our model can handle multiple visual (multiple images) and text sources (headline, main content, etc.) for a given news sample. 
\textcolor{black}{For simplification, we assume a single image and text source associated with a news sample.}
} To this end, we propose an end-to-end trainable model, \textbf{GAME-ON}: (\textbf{G}raph \textbf{A}ttention Network-based \textbf{M}ultimodal Fusi\textbf{ON}) (see Figure ~\ref{fig:model_diagram}). Our proposed framework consists of four stages to overcome the inherent challenges caused by heterogeneous modalities:

\begin{enumerate}
    \item represent a news sample (${N_i = \{T_i, V_i\}}$) into two fully connected graphs with nodes extracted from the text (${G_T}$) and visual (${G_V}$) components respectively (subsection \ref{subsec:feature-encoders}),
    \item learn common space representations for the different modalities, and add inter-modal connections between text and visual nodes (${G_{MM} = G_T + G_V}$) (subsection \ref{subsec:multimodal-space-graph}),
    \item utilize graph attention layer on the multimodal graph to weigh inter and intra-modal relations dynamically, to learn robust multimodal representation (subsection \ref{subsec:graph-layers})
    \item pooling and classification (subsection \ref{subsec:fake-news-detector})
\end{enumerate}

\textcolor{black}{The following subsections are categorized as follows. Section  \ref{subsec:feature-encoders} describes the encoding structure of the two modalities, text and image; Section \ref{subsec:multimodal-space-graph} describes how the architecture learns the common space representations for each modality; Section \ref{subsec:graph-layers} discusses the weightage given to inter and intra-modal interactions. Finally, Section \ref{subsec:fake-news-detector} determines whether or not the news sample is fake.}

\subsection{Visual and Textual Feature Encoders}\label{subsec:feature-encoders}

\subsubsection{Visual Feature Encoder} 

To this module, we input image \textit{V} accompanying the news sample and aim to extract both global and localized representations to better capture the semantics and fine-grained structure of the image. We employ a pre-trained \textit{Faster R-CNN} \cite{ren2015faster} model to extract bounding boxes of objects present in the image ${V = \{bb_1, bb_2, ..., bb_l\}}$, where ${bb_i}$ is the ${i^{th}}$ object detected by the model in image \textit{V} (global representation of the whole image is represented as ${bb_0}$, which is the uncropped image). We then use these cropped bounding boxes and the whole image as input to a pre-trained Resnet-50 \cite{he2016deep} model,  our vision model (VM), to extract local and global feature embeddings for the image. The output representation from the vision  model ${\tilde{v}_i}$ is given by  

\begin{equation}
    \tilde{v}_i =  VM(bb_i),  \forall   b_i \subset \{bb_0, bb_1, bb_2, ..., bb_l\}
\end{equation}
where ${\tilde{v}_i}$ is a 2048-dimensional vector. 
\rev{The ${\tilde{v}_i}$ feature is then resized to a 768-dimensional feature vector ${v_i}$ to match with the textual feature dimensions. Keeping in line with our simplistic and effective approach, we use mean pooling to achieve the required size. }

For the visual modality graph ${G_V = (N_V, E_V)}$, the nodes ${N_V}$ are represented by the extracted visual features (embeddings) from the vision  model ${N_V = \{v_0, v_1, v_2, ... , v_l\}}$. We assume that there exists a relationship between each visual node \cite{jiang2020kbgn}. Therefore, the visual graph is fully connected with unweighted and bi-directional edges.

\subsubsection{Textual Feature Encoder} 

To this module, we input the text accompanying the news sample, which can be viewed as a sequential list of tokens or words ${T = \{w_1, w_2, ... w_k\}}$. To capture the semantics and syntactic-level representations of the text (following the same encoding strategy as the visual encoder), we extract both context-aware token-wise representations, along with a cumulative text representation from the input text by utilizing a pre-trained Bidirectional Encoder Representations from Transformers (\textit{BERT}) \cite{devlin2018bert} as our language model (LM). Specifically, we employ the [CLS] and token-wise output embeddings for cumulative and token-level feature embeddings, respectively. The output contextualized representation ${t_i}$ from the language model is given by 

\begin{equation}
    \{t_0, t_1, t_2, ... , t_k\} =  LM({[CLS], w_1, w_2, ... w_k}),
\end{equation}
where ${t_i}$ is a 768-dimensional feature vector, ${t_0}$ is the text-level representation.

For the textual modality graph ${G_T = (N_T, E_T)}$, the nodes are represented by the extracted textual features from the language model ${N_T = \{t_0, t_t, t_2, ... , t_k\}}$. Similar to the visual graph, we assume a relationship between each textual node for consistency. Therefore, the textual graph is also fully connected with unweighted and bi-directional edges. 
\shakshi{As we are using transformer based pre-trained model for extracting features for the textual nodes, we follow the same steps of tokenization and pre-processing used in those model. We don't do any external filtering of stop words, or lemmatization, so as to not interfere with feature representation of the text, and also let the dynamic graphical attention layer to decide the important node in the text. If during training, the nodes corresponding to tokens of stopwords are not found useful, they would be downvoted. However, to not restrict our model, we refrain from doing any filtering to avoid information loss.}

\subsection{Shared Multimodal Space and Multimodal  Graph Construction}\label{subsec:multimodal-space-graph}

\subsubsection{Shared Multimodal Space} 

\textcolor{black}{In previous multimodal representation learning and information retrieval works, researchers have discussed the need to bridge the \textit{heterogeneity gap} which is created due to the presence of different modalities and their specialized pre-trained feature encoders \cite{peng2019cm}. To fully capture the semantic correlation between the heterogeneous modalities, a shallow fully connected network or a weight-sharing network is employed to learn meaningful multimodal representations in a shared feature space \cite{peng2016cross}. Taking inspiration from these works, we also utilize a two layer fully-connected feed-forward layer with dropout and ReLU non-linearity, to bridge the heterogeneity gap between the modalities and project the node features of textual ${m^t_j}$ and visual ${m^v_i}$ uni-modal graphs to common embedding space.}
\begin{equation}
    m^v_i = fully\_connected\_network(v_i),  \forall  v_i \subset \{v_0, v_1, ..., v_k\} 
\end{equation} 
\begin{equation}
    m^t_j = fully\_connected\_network(t_j),  \forall t_j \subset \{t_0, t_1, ..., t_l\}
\end{equation}
where, 
$m^v_i$ and $m^t_j$ are visual and textual feature representations.
\textcolor{black}{{A shared space for projecting multimodal embeddings has been widely used in literature. Such a weight-sharing layer becomes necessary in the age of pre-trained models, as the output distribution of two different models can be very different. By employing this shared space module, we are able to project these heterogeneous distributions or representations to a common space for better joint representation learning to bridge the heterogeneity gap.}}

\subsubsection{Multimodal Graph Construction}
To simultaneously model the inter- and intra-modal semantic relationship between the heterogeneous modalities, we introduce unweighted and bidirected edges between them to bridge the heterogeneity gap even further. Therefore, we connect each text node to every image node and vice-versa. Classical concatenation or cross-attention methods can only model cross-modal relationships from a global standpoint; however, these inter-modal and intra-modal connecting edges help the model to learn from dependencies arising from both within and across modalities concurrently at a more granular level. Given the individual graphs ${G_T}$ and ${G_V}$, we construct a multimodal graph ${G_{MM} = (N_{MM}, E_{MM})}$, with the nodes 
\begin{equation}
    N_{MM} = \{\tilde{m}^0_0, \tilde{m}^0_1, ..., \tilde{m}^0_{k+l+1}\} = \{m^v_0, ..., m^v_k, m^t_0, ..., m^t_l\},
\end{equation}
 for simplicity we are taking ${\tilde{m}^0_i}$ as the combined notation for the textual ${m^t_j}$ and visual ${m^v_i}$ nodes in our multimodal graph, and 
 
 \begin{equation}
     E_{MM} = \{e_{ij}\}^{(k+l+2) \times (k+l+2)},
 \end{equation}
matrix for the fully connected multimodal graph.

\subsubsection{Early Fusion of Modalities}
At the start of our multimodal fusion pipeline, we fuse the two modalities by projecting them into a shared space and creating multimodal edges to let the flow of information amongst these heterogeneous nodes. 
In the previous work \cite{wang2022instance} the authors encode features by two separate Multi-Layer Perceptron (MLP) blocks which does help them match the feature dimensionality for both the modalties, but doesn't bridge the heterogeneous gap by projecting them to common space.
The fact that the modalities are able to interact with each other before any trainable layer or fine-tuned network, we refer to this as \textit{early fusion}. 
Most of the previous studies \cite{wang2018eann,khattar2019mvae,qian2021hierarchical,wang2022instance} train or fine-tune specialized single-modality encoders and extra fully connected layers on the dataset in question before concatenating or fusing the modalities at a later stage in the architecture, that is, after the above-mentioned trainable layers and parameters (\textit{late fusion}). Whereas, GAME-ON helps in removing these extraneous parameters and thereby significantly reducing the number of trainable parameters involved by focusing on a multimodal representation learning from the first layer of the model to help minimize the heterogeneity gap present between modalities. We also empirically show how fusing the modalities at a later stage in our pipeline has a degrading effect on the performance of our model (Section \ref{subsec:ablation-study}).

\subsection{Graph Attention Layer}\label{subsec:graph-layers}
To learn discriminative and coherent representation for each node in our multimodal graph, we employ Graph Attention Network (\textit{GAT}) \cite{velivckovic2017graph} layer. The \textit{GAT} layer allows nodes to adaptively select between inter- and intra-modal connections concurrently and attend to more semantically relevant nodes in the fully connected graph. Unlike classical methods that employ separate blocks for inter- and intra-modal fusion, \textit{GAT}-based layers help assign different weights to multimodal neighborhood nodes, allowing robust multimodal feature representation at a more granular level. The node embedding ${\tilde{m}^{l}}$ update equation for the ${l^{th}}$ layer is given by:  
\begin{equation}
    \tilde{m}^{l} = \sum_{j\in \mathcal{N}(i)} \alpha_{i,j} W^{l} \tilde{m}_j^{l-1}
\end{equation}      
where ${\alpha_{ij}}$ is the attention score between node \textit{i} and node \textit{j} given by 
\begin{equation}
    \alpha_{ij}^{l} = \mathrm{softmax_i} (e_{ij}^{l})
\end{equation}
\begin{equation}
    e_{ij}^{l} = \mathrm{LeakyReLU}\left(\vec{a}^T [W \tilde{m}_{i} \| W \tilde{m}_{j}]\right)
\end{equation}
where ${\|}$ is the concatenation operation.

\noindent \textbf{Motivation behind GAT-Layer and fully-connected multimodal graph:} As mentioned above, creating a fully-connected graph with both intra- and inter-modal edges, along with the GAT layer to weigh the contribution of different types of neighbors according to the supervised task, helps substantially reduce the number of parameters required to learn robust multimodal features and relationships. It also helps reduce human bias from entering into the multimedia representation learning phase by neither restricting the number of nodes present in a multimodal data point nor hardcoding relationships amongst multimodal nodes to create a sparse graph. Previously, researchers have utilized separate co-attention and cross-attention modules to learn global multimodal representations by using specialized transformer layers, which unnecessarily increased the model complexity. Therefore, in GAME-ON, we learn semantically rich representations by employing GAT on a fully-connected multimodal graph to restrict the number of parameters required in the end-to-end pipeline. We also empirically show how replacing the GAT layer with a normal GCN layer in our pipeline has a degrading effect on the performance of our model  for the task of fake news detection in the ablation studies (Section \ref{subsec:ablation-study}).

\subsection{Fake News Detector}\label{subsec:fake-news-detector}

\subsubsection{Pooling Operator} 
Following our efficient and straightforward architectural choices, we use global max node pooling for selecting a graph-level representation (rather than modality specific features, and then concatenating them at the end which defeats the purpose of learning powerful multimodal homogeneous representation for each node) for our multimodal news-level graph: 
\begin{equation}
  h_{G_{MM}} = pool(\{\tilde{m}^l_0, \tilde{m}^l_1, ..., \tilde{m}^l_{k+l+1}\}) 
\end{equation}

\subsubsection{Classification} 
To this module, we input the learned multimodal graph-level representation and aim to classify the news sample as fake or real. We employ a shallow two-layer fully connected feed-forward network with intermediate ReLU non-linearity and output dimension to match with target space of two classes followed by a softmax function to convert logits into class probabilities. 
\textcolor{black}{Let ${\hat{P}_n = [\hat{P}^0_n, \hat{P}^1_n]}$ be the output predicted vector and ${Y_n}$ be the ground-truth label for the ${n^{th}}$ news sample, where ${\hat{P}^0_n}$ and ${\hat{P}^1_n}$ are the predicted probability of the ${n^{th}}$ sample being real and fake, respectively. }
We minimize the cross-entropy loss to train the model.
 \begin{equation}
     L(\theta)=  \frac{1}{N} \sum_{i=1}^{N} Y_i \log \hat{P}^0_i + (1-Y_i) \log (1-\hat{P}^1_i)
 \end{equation}

where \textit{N} is the number of news samples.

\section{Experiments and Results}\label{sec:experiments}

\subsection{Datasets and Implementation Details}\label{subsec:Datasets}

We evaluate the efficacy of the proposed GAME-ON framework on two widely adopted social media multimodal fake news datasets:
(i) MediaEval (Twitter) \cite{boididou2015verifying} and (ii) Weibo \cite{jin2017multimodal}. 
\textcolor{black}{Table ~\ref{tab:data-stats} shows the statistics of the two datasets.}

\begin{table}[]
\centering
\begin{tabular}{|c|c|c|}
\hline
                & Twitter & Weibo \\ \hline
\# of Real News & 6,026    & 4,749  \\ \hline
\# of Fake News & 7,898    & 4,779  \\ \hline
\# of Images    & 514     & 9,528   \\ \hline
\end{tabular}
\caption{Statistics of two real-world datasets.}
\label{tab:data-stats}
\end{table}

The \textbf{Twitter dataset} was released for the Verifying Multimedia Use Task as a part of MediaEval. 
\rev{There are around ten events included in this dataset such as Hurricane Sandy, Boston Marathon bombings, Columbian Chemicals Plant explosion, Nepal 2015 earthquakes, Solar Eclipse, etc. More information can be found on the MediaEval Challenge Website. \footnote{http://www.multimediaeval.org/mediaeval2015/verifyingmultimediause/}}
It consists of two parts: the development and the test set. For fair evaluation with other baselines, we use the development set for training and the test set for the evaluation. 

Each data sample (tweet) in the Twitter dataset consists of 3 sources: tweet text, associated image or video, and the additional social context with the tweet. We remove the tweets without any text or images. The dataset contains 6,026 real, 7,898 fake tweets, and 514 images (here multiple tweets are associated with a single image). For fair evaluation with other baselines, we keep the same data split scheme.

It's important to keep in mind that the dataset size may appear small simply by looking at the number of tweets (6,026 real tweets and 7,898 fake tweets). However, the emphasis of this paper is on the validation of the usage of multimedia in tweets, which necessitates combining the analysis of the textual content, and related images. As a result, the dataset's size is justified by the task's complexity and the requirement for detailed analysis of several modalities.
The dataset also contains 514 images, and it is noted that a single image may be linked to several tweets. This suggests that the dataset includes many examples of tweets utilizing multimedia, providing a wide variety of samples from which to train and assess the models.

\textcolor{black}{
The \textbf{Weibo dataset} was collected from a Chinese microblogging website, Sina Weibo \cite{jin2017multimodal}. We used the publicly available version of this dataset used by the authors of \cite{wang2018eann}. 
During pre-processing, they removed low-quality and duplicate images to ensure the overall standard of the dataset and posts without multimodal content.
The fake posts in the dataset were verified on the official rumor debunking system of Weibo. The real posts were verified by an authoritative news agency in China, Xinhua News Agency.
The dataset contains 4,749 real, 4,779 fake tweets, and 9,528 images.
}

As this dataset comes from a Chinese platform, it differs from the Twitter dataset in terms of source of data and perspective. This enables cross-cultural analysis and evaluation of multimedia utilizing the same framework.

Overall, the selection and volume of these datasets were most likely driven by their research problem's suitability, availability, and the goal of providing a broad range of examples for training and testing the proposed framework.

\subsection{Implementation Details}\label{subsec:Implementation Details}
We implemented our proposed GAME-ON framework via Deep Graph Library\footnote{https://github.com/dmlc/dgl}. 
\textcolor{black}{and PyTorch\footnote{https://github.com/pytorch/pytorch} framework. For all the experiments, we use GeForce GTX 1080 Ti GPU with 11 GB memory. 
We use the pre-trained `bert-base-uncased' and `bert-base-chinese' 
BERT models 
from HuggingFace\footnote{https://github.com/huggingface/transformers} and pre-trained  CNN model 
from Torchvision\footnote{https://pytorch.org/vision/stable/models.html} }
for textual and visual node feature extraction, respectively. The common embedding space dimension to which we project the text and image features is 512 with 768 as the intermediate dimension. We used a \rev{ single GAT layer with output dimension as 256}, with one attention head. For the fake news detector, we use an intermediate feature dimension of 128 and 64, 
0.5 dropout rate. We used an effective batch size of 512 and trained the model end-to-end with Adam optimizer \cite{kingma2014adam}, with an initial learning rate of 1e-4 with a linearly decreasing scheduler. \rev{For the purpose of reproducibility, we have released our code, and the random seeds, random seed = 5, for our experiments to replicate the results.\footnote{https://github.com/mudit-dhawan/GAME-ON}}

\subsection{Multimodal Baselines}\label{subsec:Multimodal Baselines}

We compare multimodal methods to assess the performance of GAME-ON. 

\noindent\textbf{EANN} \cite{wang2018eann} extracts textual and visual data separately using pre-trained models and combines them to feed into a fake news detector and event discriminator to eliminate any event-specific features.

\noindent\textbf{MVAE} \cite{khattar2019mvae} uses a bimodal variational autoencoder and a fake news classifier, this model provided a shared representation of multimodal data.

\noindent\textbf{SAFE} \cite{zhou2020safe} uses a modified cosine similarity to encode the relevance between textual and visual information.

\noindent\textbf{CALM} \cite{wu2021cross} employs a cross-modal fusion network with orthogonal latent memory, the CALM framework is utilized to identify rumors. 

\noindent\textbf{MCAN} \cite{wu2021multimodal} uses various co-attention layers are used to learn inter-modality relations, with visual features being fused first, followed by textual features.

\noindent\textbf{DIFF} \cite{wang2022instance} creates a multimodal fully connected graph, and sequentially performs intra-modal feature aggregation and inter-modality fusion with features in two different embedding spaces. 

\textcolor{black}{
\noindent\textbf{CSFND} \cite{peng2024not} efficiently combines context information into representation learning. It utilizes contextual testing techniques and unsupervised context learning.
}

\subsection{Results}
Table ~\ref{tab:baseline-comparison-results} 
demonstrates the experimental results of our proposed model GAME-ON on the two datasets and the baselines.

\subsubsection{Performance comparison} 

\textcolor{black}{
The accuracy, precision, recall, and F1 score performance metrics for the various approaches on the Weibo and Twitter datasets are shown in Table \ref{tab:baseline-comparison-results}. Notably, GAME-ON obtains an accuracy of 95.8\% on the Twitter dataset, with solid performance for recognizing real news and very good recall (91.2\%) and precision (97.2\%) for fake news identification as well as strong results in detecting real news.
The reason for the GAME-ON model's lower accuracy of 88.9\% on the Weibo dataset is due to the fact that Weibo dataset is not the same as the Twitter dataset due to differences in language usage, cultural contexts, and shared content categories. These variations may provide particular difficulties for algorithms used to detect fake news, which may have an effect on performance indicators as a whole. Despite these challenges, the GAME-ON model's competitive precision, recall, and F1 scores on the Weibo dataset highlight its resilience and efficiency in detecting fake news.
}

In a nutshell, we observe that the GAME-ON model surpasses the current state-of-the-art model CALM on the Twitter dataset by an average of \textbf{11\%} in all the performance metrics. On the Weibo, GAME-ON achieves competitive performance in comparison to the complex and large state-of-the-art models while having significantly less number of trainable parameters.
\textcolor{black}{Even though the Weibo dataset's decreased accuracy may be notable, it has no bearing on the GAME-ON framework's overall efficacy and relevance because it still performs well on different social media platforms, namely, Twitter and Weibo.}

\begin{table}[!htbp]
\centering
\resizebox{\textwidth}{!}{%
\begin{tabular}{|c|c|c|ccc|ccc|}
\hline
\multirow{2}{*}{\textbf{Dataset}} & \multirow{2}{*}{\textbf{Method}} & \multirow{2}{*}{\textbf{Accuracy}} & \multicolumn{3}{c|}{\textbf{Fake News}}                                                         & \multicolumn{3}{c|}{\textbf{Real News}}                                                         \\ \cline{4-9} 
                                  &                                  &                                    & \multicolumn{1}{c|}{\textbf{Precision}} & \multicolumn{1}{c|}{\textbf{Recall}} & \textbf{F1}    & \multicolumn{1}{c|}{\textbf{Precision}} & \multicolumn{1}{c|}{\textbf{Recall}} & \textbf{F1}    \\ \hline
\multirow{7}{*}{\textbf{Twitter}} & EANN                             & 0.648                              & \multicolumn{1}{c|}{0.810}              & \multicolumn{1}{c|}{0.498}           & 0.617          & \multicolumn{1}{c|}{0.584}              & \multicolumn{1}{c|}{0.759}           & 0.660          \\ \cline{2-9} 
                                  & MVAE                             & 0.745                              & \multicolumn{1}{c|}{0.801}              & \multicolumn{1}{c|}{0.719}           & 0.7558         & \multicolumn{1}{c|}{0.689}              & \multicolumn{1}{c|}{0.777}           & 0.730          \\ \cline{2-9} 
                                  & CALM                             & 0.845                              & \multicolumn{1}{c|}{0.785}              & \multicolumn{1}{c|}{0.831}           & 0.807          & \multicolumn{1}{c|}{0.888}              & \multicolumn{1}{c|}{0.855}           & 0.871          \\ \cline{2-9} 
                                  & SAFE                             & 0.766                              & \multicolumn{1}{c|}{0.777}              & \multicolumn{1}{c|}{0.795}           & 0.786          & \multicolumn{1}{c|}{0.752}              & \multicolumn{1}{c|}{0.731}           & 0.742          \\ \cline{2-9} 
                                  & MCAN                             & 0.809                              & \multicolumn{1}{c|}{0.889}              & \multicolumn{1}{c|}{0.765}           & 0.822          & \multicolumn{1}{c|}{0.732}              & \multicolumn{1}{c|}{0.871}           & 0.795          \\ \cline{2-9} 
                                                     & CSFND                             &  0.833                              & \multicolumn{1}{c|}{ 0.899}              & \multicolumn{1}{c|}{0.799}           & 0.846          & \multicolumn{1}{c|}{0.763}              & \multicolumn{1}{c|}{0.878}           & 0.817          \\ \cline{2-9} 
                                  & DIIF                             & 0.783                              & \multicolumn{1}{c|}{0.810}              & \multicolumn{1}{c|}{0.803}           & 0.806          & \multicolumn{1}{c|}{0.758}              & \multicolumn{1}{c|}{0.786}           & 0.772          \\ \cline{2-9} 
                                  & GAME-ON                          & \textbf{0.958}                     & \multicolumn{1}{c|}{\textbf{0.972}}     & \multicolumn{1}{c|}{\textbf{0.912}}  & \textbf{0.942} & \multicolumn{1}{c|}{\textbf{0.926}}     & \multicolumn{1}{c|}{\textbf{0.901}}  & \textbf{0.913} \\ \hline
\multirow{7}{*}{\textbf{Weibo}}   & EANN                             & 0.782                              & \multicolumn{1}{c|}{0.827}              & \multicolumn{1}{c|}{0.697}           & 0.756          & \multicolumn{1}{c|}{0.752}              & \multicolumn{1}{c|}{0.863}           & 0.804          \\ \cline{2-9} 
                                  & MVAE                             & 0.824                              & \multicolumn{1}{c|}{0.854}              & \multicolumn{1}{c|}{0.769}           & 0.809          & \multicolumn{1}{c|}{0.802}              & \multicolumn{1}{c|}{0.875}           & 0.837          \\ \cline{2-9} 
                                  & CALM                             & 0.846                              & \multicolumn{1}{c|}{0.843}              & \multicolumn{1}{c|}{0.864}           & 0.853          & \multicolumn{1}{c|}{0.851}              & \multicolumn{1}{c|}{0.828}           & 0.839          \\ \cline{2-9} 
                                  & SAFE                             & 0.763                              & \multicolumn{1}{c|}{0.833}              & \multicolumn{1}{c|}{0.659}           & 0.736          & \multicolumn{1}{c|}{0.717}              & \multicolumn{1}{c|}{0.868}           & 0.785          \\ \cline{2-9} 
                                  & MCAN                             & \textbf{0.899}                     & \multicolumn{1}{c|}{\textbf{0.913}}     & \multicolumn{1}{c|}{0.889}           & 0.901          & \multicolumn{1}{c|}{0.884}              & \multicolumn{1}{c|}{\textbf{0.909}}  & \textbf{0.897} \\ \cline{2-9} 
                                     & CSFND                             & 0.895                     & \multicolumn{1}{c|}{0.899}     & \multicolumn{1}{c|}{0.894}           &  0.897         & \multicolumn{1}{c|}{0.892}              & \multicolumn{1}{c|}{0.894}  & 0.894 \\ \cline{2-9} 
                                  & DIIF                             & 0.890                              & \multicolumn{1}{c|}{0.912}              & \multicolumn{1}{c|}{\textbf{0.894}}  & \textbf{0.903} & \multicolumn{1}{c|}{\textbf{0.905}}     & \multicolumn{1}{c|}{0.879}           & 0.892          \\ \cline{2-9} 
                                  & \textbf{GAME-ON}                 & \textit{0.889}                     & \multicolumn{1}{c|}{\textit{0.891}}     & \multicolumn{1}{c|}{\textit{0.890}}  & \textit{0.890} & \multicolumn{1}{c|}{\textit{0.887}}     & \multicolumn{1}{c|}{\textit{0.886}}  & \textit{0.886} \\ \hline
\end{tabular}%
}
\caption{Performance comparison of GAME-ON framework with state-of-the-art baselines on Twitter and Weibo datasets. GAME-ON outperforms the baselines on Twitter, and achieves comparable performance on Weibo.
}
\label{tab:baseline-comparison-results}
\end{table}

\subsubsection{Model size} 
We also compare our proposed framework with the comparable baselines on the axis of the model's size or the number of trainable parameters \textcolor{black}{ (Table \ref{tab:model-size-results})}. 
Our model contains the least number of trainable parameters when compared to other state-of-the-art models with comparable performance. The closest baseline model, DIFF, has roughly ~6.6 million trainable parameters, whereas our model contains around 599,362 trainable parameters, which is $\sim$\textbf{91\%} less than the best comparable baseline. Thus, \textit{GAME-ON} is more feasible for deployment in real-world applications.

\begin{table}[!htbp]
\centering
\begin{tabular}{|c|c|}
\hline
\textbf{Model Name}       & \textbf{\# of Trainable Parameters} \\ \hline
SpotFake         &  $\sim$ 11 million +               \\ \hline
 CALM             &  9,954,888                \\ \hline
HMCAN            &  $\sim$ 20 million +               \\ \hline
MCAN             &  $\sim$ 2.9 million +                \\ \hline
\textbf{GAME-ON} &  \textbf{1,017,730}   ($\sim$ 1 million)             \\ \hline
 \end{tabular}
 \caption{Comparison of GAME-ON with baselines based on model size (\# of trainable parameters). GAME-ON utilizes 65\% fewer trainable parameters than MCAN.}
\label{tab:model-size-results}
\end{table}
\rev{
\subsection{Ablation Study} \label{subsec:ablation-study}

To study the importance of our fusion framework, we conducted an ablation study, and the results are summarized in Table \ref{tab:ablation-results}.
Specifically, we compare our GAME-ON model with unimodal graphical approaches (Textual and Visual in Table \ref{tab:ablation-results}) to demonstrate the importance of our multimodal framework. We also experiment to understand the significance of adaptive selection (using graph attention layer) of important multimodal nodes in the graph for a given news sample. 

Other than the lightweight attention based graphical interaction, we also look at the importance of early fusion, and inter-modality interaction. For this we once remove the inter-modal edges from the graph and separately train to different graphs for the modalities, and rather than combining them early in the pipeline, we concatenate them at a later stage to shed some light on the importance granular intermodal interaction and early fusion.
We compare our framework with two other multimodal graphical methods to evaluate the efficacy of two significant improvements in our proposed model:

\noindent\textbf{1. Absence of Graph Attention Layer in a multimodal fully-connected graph (GCN-Fusion).} The GAT layer attends to relevant nodes in the fully connected graph adaptively and simultaneously weighs multimodal connections for granular inter-and intra-modal interactions to learn the optimal contribution of both. To this end, we replace our GAT layer with a simple graph convolution (GCN) layer to highlight its importance. GAME-ON model outperforms this GCN based model (\textbf{GCN-Fusion} model in Table \ref{tab:ablation-results}) on both datasets by an average of \textbf{3.6\%} and \textbf{3.3\%} in F1-score and accuracy, respectively.

\begin{table}[!htbp]
\centering
\begin{tabular}{|c|c|cc|cc|}
\hline
                                     & \textbf{Dataset} & \multicolumn{2}{c|}{\textbf{Twitter}}                                 & \multicolumn{2}{c|}{\textbf{Weibo}}                                      \\ \hline
\textbf{Type}                        & \textbf{Method}  & \multicolumn{1}{c|}{\textbf{Accuracy}}       & \textbf{F1-Score}               & \multicolumn{1}{c|}{\textbf{Accuracy}} & \textbf{F1-Score}               \\ \hline
\multirow{2}{*}{\textbf{Unimodal}}   & Textual                           & \multicolumn{1}{c|}{0.558}          & 0.508                           & \multicolumn{1}{c|}{0.820}             & 0.822                           \\ \cline{2-6} 
                                     & Visual                            & \multicolumn{1}{c|}{0.862}          & 0.677                           & \multicolumn{1}{c|}{0.537}             & 0.556                           \\ \hline
\multirow{3}{*}{\textbf{Multimodal}} & Concatenation                     & \multicolumn{1}{c|}{0.897}          & 0.782                           & \multicolumn{1}{c|}{0.840}             & 0.841                           \\ \cline{2-6} 
                                     & GCN-Fusion                        & \multicolumn{1}{c|}{0.929}          & 0.900                           & \multicolumn{1}{c|}{0.846}             & 0.850                           \\ \cline{2-6} 
                                     & \textbf{GAME-ON} & \multicolumn{1}{c|}{\textbf{0.958}} & \textbf{0.924} & \multicolumn{1}{c|}{\textbf{0.889}}    & \textbf{0.892} \\ \hline 
\end{tabular}
\caption{Performance comparison of GAME-ON framework with its variants on Twitter and Weibo datasets. GAME-ON outperforms its unimodal and multimodal variants on both datasets.}
\label{tab:ablation-results}
\end{table}

\noindent\textbf{2. Absence of Granular inter-modal interaction and early fusion (Concatenation).} 
\textcolor{black}{
Unlike previous multimodal fake news works, in GAME-ON, we propose to fuse the modalities at an early stage (before trainable parameters or fine-tuned specialized encoders) in the pipeline. 
It leads to a novel fine-grained representation of learning from dependencies occurring within and across modalities concurrently.
We train a late-fusion model without allowing the fine-grained inter-modal flow of information (\textbf{Concatenation} model in Table \ref{tab:ablation-results}) to emphasize that fusion of text and image information at a later stage will cause information loss and the significance of granular inter-modal interactions. In this model, we remove the inter-modal connections and feed textual and visual unimodal graphs to the GAT layer individually, followed by mean pooling and concatenating the global visual and textual unimodal graphical representations. Results in Table \ref{tab:ablation-results} indicate that introducing inter-modal edges and allowing the model to learn both intra- and inter-modal dependencies simultaneously decreases the heterogeneity gap and improves the performance on both datasets by an average of \textbf{5.5\%} and \textbf{6.75\%} in accuracy and F1-score, respectively.
}
\textcolor{black}{The primary objective of our research is to improve model performance by combining two different modalities. For assessing individual modality performance, the unimodal (visual) approach performs better than the unimodal (textual) approach which can be attributed to the richer information acquired by images that comprise several interrelated objects.}

}

Inspired by their limitation of inefficient fusion of different modalities with complex models and the recent success of GNNs \cite{wang2022instance}, we propose, GAME-ON, an end-to-end trainable GNN based framework. The GAME-ON framework contains fewer parameters but also takes care of the heterogeneity gap. 

\section{Conclusion}\label{sec:conclusion}

In this work, we proposed GAME-ON, an end-to-end trainable GNN based framework for detecting fake news. We evaluated our framework's efficacy on two publicly available multimodal datasets; Twitter and Weibo. 
Our work overcomes two significant shortcomings of previous works. First, works that encode the inter-modal relation using the concatenation operator fail to address the heterogeneity gap in multimodal data. Second, even research that addressed the first problem used complicated models to combine the various modalities, implying a higher risk of overfitting.
\textcolor{black} {In order to overcome the limitations of existing approaches, our framework transforms a multimodal data point into a multimodal graphical structure with objects and words represented as nodes and edges that enable granular interactions within and across distinct modalities to merge them early in the framework. }
Our model outperforms the current state-of-the-art models on the Twitter by an average of \textbf{11\%}, and on Weibo maintains a competitive performance.
In addition, we compared our model with state-of-the-art models in terms of trainable parameters, and our model has $\sim$\textbf{91\%} fewer parameters than the best comparable baseline. Therefore, \textit{GAME-ON} is more feasible for deployment in real-world applications.
\textcolor{black} {It is to be noted that the proposed framework is not restricted to the problem of fake news detection only. Instead, this approach can be applied to various tasks that involve multimodal or heterogeneous data types and their fusion. For example, it can be applied to multimodal sentiment classification, toxicity detection, and hateful memes prediction. In addition, our approach addresses the critical issue of inefficient fusion of multiple modalities and simultaneous encoding of intra- and inter-modal relations using significantly less parameters. Therefore, \textit{GAME-ON} can be deployed in production systems which pose strict constraint on working memory and latency to detect and curb fake news on multimodal social media platforms} 
For future work, this framework can be extended to include datasets with long articles. 

\section{Statements and Declarations}
\subsection{Funding}
R.Sharma's work has received funding from the EU H2020 program under the SoBigData++ project (grant agreement No. 871042), by the CHIST-ERA grant CHIST-ERA-19-XAI-010,  ETAg (grant No. SLTAT21096), and partially funded by HOLISTIC ANALYSIS OF ORGANISED MISINFORMATION ACTIVITY IN SOCIAL NETWORKS project (PCI2022-135026-2).

\subsection{Availability of data and material}
The authors used widely popular and publicly available multimodal fake news datasets. For more details, please refer to Section 4.1 (Datasets and Implementation Details).

\subsection{Code availability}
The code repository is available at https://github.com/mudit-dhawan/GAME-ON.

\subsection{Authors' contributions}
\textit{Mudit Dhawan:} Formal analysis, Investigation, Validation, Visualization, and Writing – original draft.

\noindent\textit{Shakshi Sharma:} Formal analysis, Investigation, Validation, Visualization, and Writing – original draft.

\noindent\textit{Aditya Kadam:} Visualization, and Writing – Review \& Editing.

\noindent\textit{Rajesh Sharma:} Supervision, and Writing – Review \& Editing.

\noindent\textit{Ponnurangam Kumaraguru:} Supervision, and Writing – Review \& Editing.

\bibliographystyle{spmpsci}
\bibliography{ref}

\end{document}